# Nanostructure-induced distortion in single-emitter microscopy

*Kangmook Lim[a,b], Chad Ropp[a], John Fourkas[c], Benjamin Shapiro[d], Edo Waks[a,b,1]*

[a]Department of Electrical and Computer Engineering and Institute for Research in Electronics and Applied Physics, University of Maryland, College Park, Maryland 20742, USA.

[b]Joint Quantum Institute, University of Maryland and the National Institute of Standards and Technology, College Park, Maryland 20742, USA.

[c]Department of Chemistry & Biochemistry and Institute for Physical Science & Technology, University of Maryland, College Park, Maryland 20742, USA.

[d]Fischell Department of Bioengineering and the Institute for Systems Research, University of Maryland, College Park, Maryland 20742, USA.

[1]To whom correspondence should be addressed. E-mail: edowaks@umd.edu.

ABSTRACT

Single-emitter microscopy has emerged as a promising method of imaging nanostructures with nanoscale resolution. This technique uses the centroid position of an emitter's far-field radiation pattern to infer its position to a precision that is far below the diffraction limit. However, nanostructures composed of high-dielectric materials such as noble metals can distort the far-field radiation pattern. Previous work has shown that these distortions can significantly degrade the imaging of the local density of states in metallic nanowires using polarization-resolved imaging. But unlike nanowires, nanoparticles do not have a well-defined axis of symmetry, which makes polarization-resolved imaging difficult to apply. Nanoparticles also exhibit a more complex range of distortions, because in addition to introducing a high dielectric surface, they also act as efficient scatterers. Thus, the distortion effects of nanoparticles in single-emitter microscopy remains poorly understood. Here we demonstrate that metallic nanoparticles can significantly distort the accuracy of single-emitter imaging at distances exceeding 300 nm. We use a single quantum dot to probe both the magnitude and the direction of the metallic nanoparticle-induced imaging distortion and show that the diffraction spot of the quantum dot can shift by more than 35 nm. The centroid position of the emitter generally shifts away from the nanoparticle position, in contradiction to the conventional wisdom that the nanoparticle is a scattering object that will pull in the diffraction spot of the emitter towards its center. These results suggest that dielectric distortion of the emission pattern dominates over scattering. We also show that by monitoring the distortion of the quantum dot diffraction spot we can obtain high-resolution spatial images of the nanoparticle, providing a new method for performing highly precise, sub-diffraction spatial imaging. These results provide a better understanding of the complex near-field coupling between emitters and nanostructures, and open up new opportunities to perform super-resolution microscopy with higher accuracy.

The diffraction of light limits the spatial resolution of conventional far-field optical microscopes to within approximately an optical wavelength[1,2]. Single-emitter imaging overcomes this limit by tracking the diffraction spots of single point emitters to a precision that is much finer than the wavelength of light[3-8]. This technique has emerged as a versatile approach for obtaining sub-wavelength information in a broad range of applications in biology[9-12] and chemistry[13-16]. More recently, a number of studies have applied single-emitter imaging techniques to probe plasmonic nanostructures such as metallic hot spots[17-19], nanowires[20,21], nanoparticles[22,23] and nanoantennas,[24,25] with reported spatial precision that is finer than 10 nm.

Single-emitter imaging of nanostructures often relies on the ability to track an emitter near a metallic or high-dielectric surface. However, such surfaces can significantly complicate the ability to track the emitter precisely[26,27]. We recently demonstrated that near the surface of a metal nanowire, an emitter induces an image dipole that significantly distorts the tracking accuracy[28]. Spherical metal nanoparticles can distort an emitter's diffraction spot in more complicated ways because it is difficult to distinguish the light scattered by the particles from the direct emission from the emitter. Because the actual position of the emitter is usually not known, the magnitude and the direction of these distortions becomes extremely difficult to measure.

In this letter, we demonstrate that nanoparticles can significantly distort the tracking accuracy of a single emitter at distances exceeding 300 nm when performing single-emitter imaging. We scan a metallic nanoparticle near an immobilized emitter (a single quantum dot) and monitor the shift in the emitter's centroid position with nanoscale accuracy. We observe a displacement of the centroid position by more than 35 nm, which is much greater than the expected accuracy of the emitter tracking algorithm[3, 29]. The centroid position of the emitter is pushed away from the surface of the nanoparticle, in contrast to the conventional wisdom that a strongly scattering metallic nanoparticle will pull the emitter's diffraction spot closer to the surface. We compare these results to full-wave finite-difference time-domain (FDTD) calculations which agree well with measurement results and corroborate the observed behavior. Finally, we demonstrate that by monitoring the distortion of the emitter's diffraction spot we can spatially image the nanoparticle with high accuracy and precision. Our results provide a better fundamental understanding of near-field coupling between emitters and nanostructures and offer a promising route towards highly accurate super-resolution imaging of nanostructures.

Figure 1a illustrates the measurement approach. We manipulate a single metallic nanoparticle near an emitter immobilized on a surface. We measure the centroid position of the emitter as a function of the metallic nanoparticle's position in order to observe a shift. In our experiments the emitter is a single CdSe/ZnS colloidal quantum dot (Ocean NanoTech) with the center wavelength at 620 nm and the nanoparticle is a nominally 150 nm diameter gold nanosphere (Sigma-Aldrich Co. LLC). We use microfluidic flow control to position the nanoparticle with nanoscale accuracy[30]. This technique uses a microfluidic cross-channel device to position the nanoparticle in two dimensions. We engineer the surrounding fluid to confine the nanoparticle to the same surface as the emitter and ensure that the two particles are located to within 100 nm of each other in the out-of-plane direction. We have reported the details of this microfluidic control method previously[31-33]. We also provide a detailed description of this technique, along with the specific fluid and microfluidic device properties, in the supporting information.

We track the observed position of the quantum dot and gold nanoparticle using a home-built inverted wide field microscope. We use a 532 nm laser as an excitation source for the quantum dot, and image the fluorescence at 620 nm. We image the gold nanoparticle using a halogen white-light source. In both cases, we excite and collect emitted or scattered light using a 100× oil-immersion objective with a numerical aperture of 1.45 that images the collected signal onto an electron-multiplying charge-coupled device (EMCCD) camera.

To monitor the distortion of the diffraction spot of the quantum dot induced by the gold nanoparticle, we must accurately track both objects even when they are sufficiently close that their diffraction spots overlap. We achieve this crucial requirement using a stroboscopic imaging method. We interleave the white light and excitation laser, synchronized with the camera frame rate of 10 Hz, as depicted in Figure 1b. When the excitation laser is on and the white light is off, we image the diffraction spot of the quantum dot. We subsequently turn off the excitation laser and turn on the white light to measure the position of the gold nanoparticle. A long-pass optical filter rejects the scatter of the excitation laser so that we image only the fluorescence from the quantum dot. Both the excitation laser and white light illuminate the sample for 50 ms. The two continuous images taken under different illuminations constitute a single measurement data point. We track the quantum dot and the gold nanoparticle in alternating frames and correlate the diffraction pattern of the quantum dot with the position of the gold nanoparticle.

Figures 1c,d are two consecutive images of an immobilized quantum dot and a nearby gold nanoparticle (separated by 335 nm). We focus on a 21 × 21 pixel area (~2.7 μm × 2.7 μm) around the collected signals and fit each diffraction spot to a two-dimensional Gaussian point-spread

function to determine the centroid position. We indicate the centroid positions of the quantum dot and gold nanoparticle in Figures 1c,d using cross markers in red and yellow, respectively. There is no measureable scattering signal from the nanoparticle when we image the quantum dot, nor do we observe any emission from the quantum dot when imaging the nanoparticle.

To determine the spatial accuracy of our system, we measured the position of the immobilized quantum dot in the absence of gold nanoparticle continuously for 4 minutes. Figure 2a is a scatter plot showing the accumulated centroid positions of the quantum dot (blue). Figures 2b,c are histograms of the measured quantum dot positions along the *x* and *y* coordinates respectively. Fits of the histograms to Gaussian distributions reveal a spatial precision of 8 nm along the *x* direction and 9 nm along the *y* direction. This spatial precision is limited by system vision noise, which includes a combination of camera read noise and multiplication noise, as well as the shot noise of the emitter[21].

To probe the diffraction spot distortion induced by a gold nanoparticle, we rastered the position of the gold nanoparticle over the area of a circle with a radius of 500 nm centered at the quantum dot. During such raster scans, mechanical vibrations and sample stage drift can change the position of the immobilized quantum dot. To correct for this drift we used a separate immobilized gold nanoparticle deposited in a different location on the sample surface as a position marker. By tracking this marker we could determine and compensate for the extent of the stage drift. We apply the correction method throughout the experiments including the data shown in Figure 2.

We map the displacement of the centroid position along the $x$ and $y$ directions as a function of the quantum dot position relative to the center of the nanoparticle. We define these displacements as $\Delta x = \tilde{x} - x_0$ and $\Delta y = \tilde{y} - y_0$, where $\tilde{x}$ and $\tilde{y}$ are the $x$ and $y$ coordinates of the centroid position of the quantum dot and $x_0$ and $y_0$ are the coordinates of the actual position of quantum dot measured when the gold nanoparticle is far away. Figures 3a,b are reconstructed images of $\Delta x$ and $\Delta y$ using a Gaussian-weighted spatial average of the measurement data. Figure 3c shows the total displacement of the centroid position, defined as $\Delta r = \sqrt{\Delta x^2 + \Delta y^2}$, as a function of the quantum dot position relative to the nanoparticle position. The solid black circle delineates the expected position of the nanoparticle surface with a diameter of 150 nm.

Figures 3a-c show that the measured quantum dot centroid position can shift by more than 35 nm in the presence of the gold nanoparticle. This displacement is symmetric about the center position of the gold nanoparticle, indicating that the diffraction spot always shifts away along the direction orthogonal to the nanoparticle's surface. Thus, we observe a displacement of the centroid along the $x$ axis when the quantum dot is located along the $x$ direction with respect to the nanoparticle, and a displacement of the centroid along the $y$ axis when the quantum dot is located along the $y$ direction with respect to the nanoparticle. The shift of the centroid is therefore not a result of scattering from the nanoparticle, which would cause the centroid to be pulled in. We observe a significant displacement of the centroid position even at distance of up to 300 nm away from the nanoparticle.

Figures 3d-f show the calculated displacements of the centroid position of the quantum dot using numerical finite-difference time-domain simulations (see supporting information) which correspond to the measurements in Figures 3a-c. The numerical calculations show good agreement with the measured displacements at large distances from the gold nanoparticle. The calculations also indicate that within 40 nm of the center of the gold nanoparticle, the diffraction spot shifts towards the gold nanoparticle instead of being pushed away. We attribute this effect to strong nanoparticle scattering at small distances below 10 nm. Because this effect is very short-ranged and localized only to a few nanometers from the nanoparticle surface, we do not have the spatial resolution to clearly resolve it. Therefore we do not observe it in the measured data,

Although the distortion effects we observe can degrade the precision of super-resolution imaging techniques, they may also contain useful information. One method to image the shape of the gold nanoparticle is by looking at the distortion of the emission intensity of the quantum dot. As the nanoparticle crosses over the quantum dot, we expect a reduction in intensity due to shadowing and absorption. In Figure 4a we plot the intensity of the quantum dot, measured by summing CCD pixels enclosing the diffraction spots, as a function of its position relative to the nanoparticle. In Figure 4b we show the calculated intensity. Although the intensity provides a re-constructed image of the gold nanoparticle, the spatial resolution is poor. We fit the re-constructed image to a Gaussian point-spread function and determine the full-width half-maximum of the nanoparticle to be 912 nm, much bigger than the nominal 150 nm diameter of the spherical gold nanoparticle (indicated by the black line). In addition, the measurement suffers from a low signal-to-noise ratio due to quantum dot blinking, which induces large intensity fluctuations[34].

Figure 4c shows the width of the quantum dot diffraction spot as a function of the quantum dot position. To obtain the width, we fit each diffraction spot to a Gaussian point-spread function and plot the standard deviation which is insensitive to the quantum dot blinking. Figure 4d shows the calculated results for comparison. The Gaussian standard deviation of the emitter provides a greatly improved reconstruction of the nanoparticle shape. Fitting the reconstructed image in Figure 4c to a Gaussian we determine the size of the gold nanoparticle to be 153 nm (using the full-width half-maximum), which is much closer to the expected size of the nanoparticle. We note that in our experiment, the spherical symmetry of the gold nanoparticle ensures that tumbling of the particle in the fluid does not distort the final image. For non-spherically symmetric particles one would instead immobilize the particle and scan the dot. Alternately, it may be possible to use other information, such as polarization or elongation of the diffraction spot, to obtain orientation information about nanoparticles such as rods, even when their orientation is changing in time.

In conclusion, we have demonstrated that the near-field coupling between an emitter and a nanoparticle can significantly distort the centroid positions of the emitter at distances exceeding 300 nm. We optically tracked a single immobilized emitter in the vicinity of a deterministically positioned nanoparticle and showed that the observed centroid position of the emitter can shift by more than 35 nm. We also showed that by probing the distortion of the emitter's diffraction spot, we can attain useful information that enables high precision spatial imaging of the shape of the nanoparticle. Although our measurements focused on probing spherical gold nanoparticles, we could potentially extend this technique to probe other non-spherical nanoparticles such as rods by using polarization resolved tracking[28], which could provide information about both position and orientation. Ultimately, this method could enable highly precise single-emitter tracking of a broad range of nanostructures for applications in imaging, sensing, and engineering of strong light-matter interactions.

## ASSOCIATED CONTENT

**Supporting Information**

Fluid composition, Microfluidic device and nanoparticle positioning, FDTD simulations. This material is available free of charge via the Internet at http://pubs.acs.org.

## AUTHOR INFORMATION

**Corresponding Author**

*E-mail: edowaks@umd.edu.

**Notes**

The authors declare no competing financial interest.

## ACKNOWLEDGMENT

The authors would like to acknowledge financial support from the Physics Frontier Center at the Joint Quantum Institute.

34	Shimizu KT, Neuhauser RG, Leatherdale CA, Empedocles SA, Woo WK, Bawendi MG: Blinking statistics in single semiconductor nanocrystal quantum dots. *Physical Review B* 2001;**63**:205316.
14

FIGURES

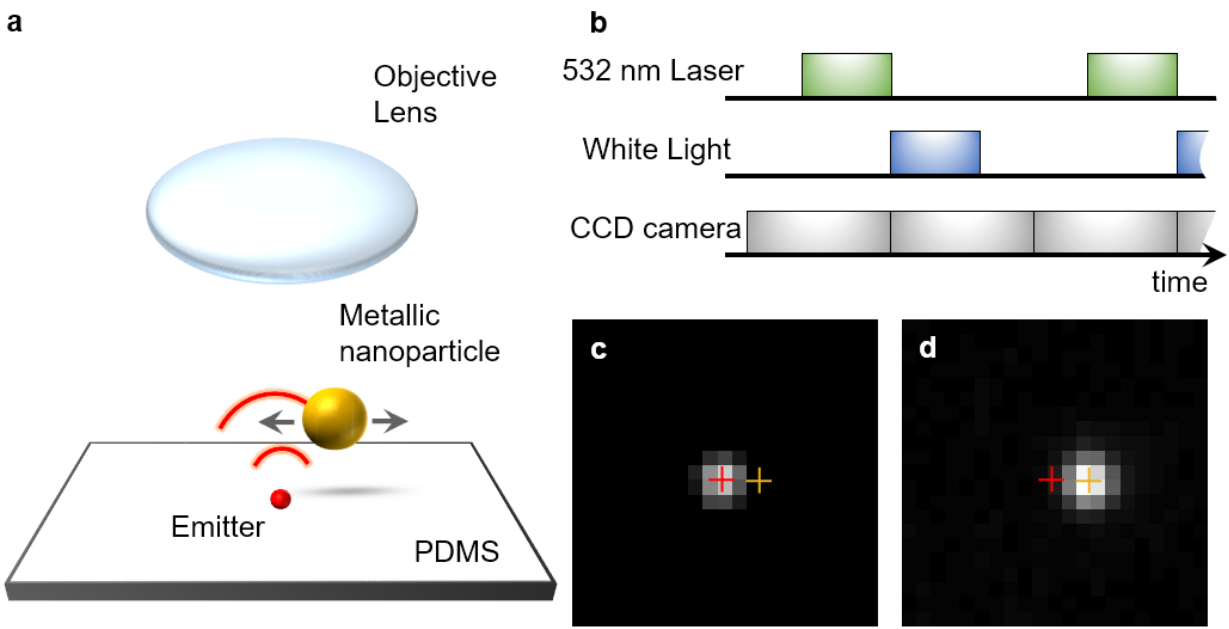

Figure 1. (a) Schematic of experimental setup. An emitter (quantum dot) is immobilized on the polydimethylsiloxane (PDMS) surface of a microfluidic device. A metallic nanoparticle (gold nanosphere) is positioned near the emitter along the PDMS surface using microfluidic flow control. An objective lens outside of the microfluidic device collects the photoluminescence signal from the emitter and the scattering signal from the nanoparticle. (b) Measurement sequence of the experiment. The CCD camera is synchronized with alternating pulses of 532 nm laser light and white light. The CCD camera exposure time is 100 ms for each frame. (c,d) Images of quantum dot and gold nanoparticle with 532 nm (c) and white light (d) illumination, respectively. The red cross marker represents the centroid position of quantum dot and yellow cross marker represents the position of gold nanoparticle. The distance between the quantum dot and the gold nanoparticle is 335 nm in these frames.

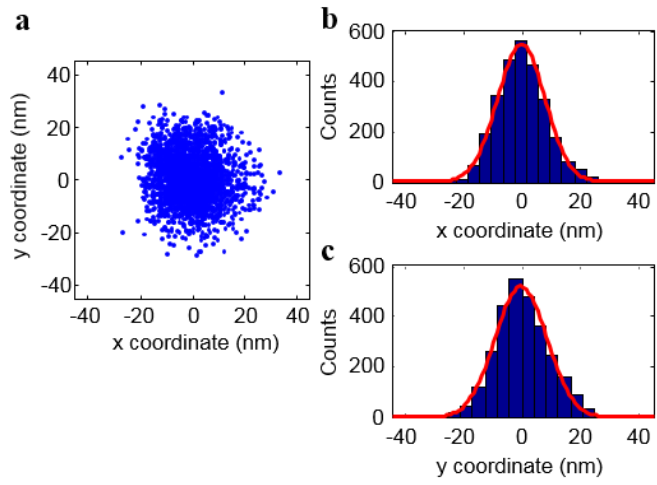

Figure 2. (a) Scatter plot of tracked positions of an isolated, immobilized quantum dot. (b) Histogram of the $x$-coordinates of the data shown in panel a. (c) Histogram of the $y$-coordinates of the data shown in panel a. For panel b-c, the measured coordinates are shown in blue and the red line is a Gaussian. The standard deviations are 8 and 9 nm in the $x$ and y axes, respectively.

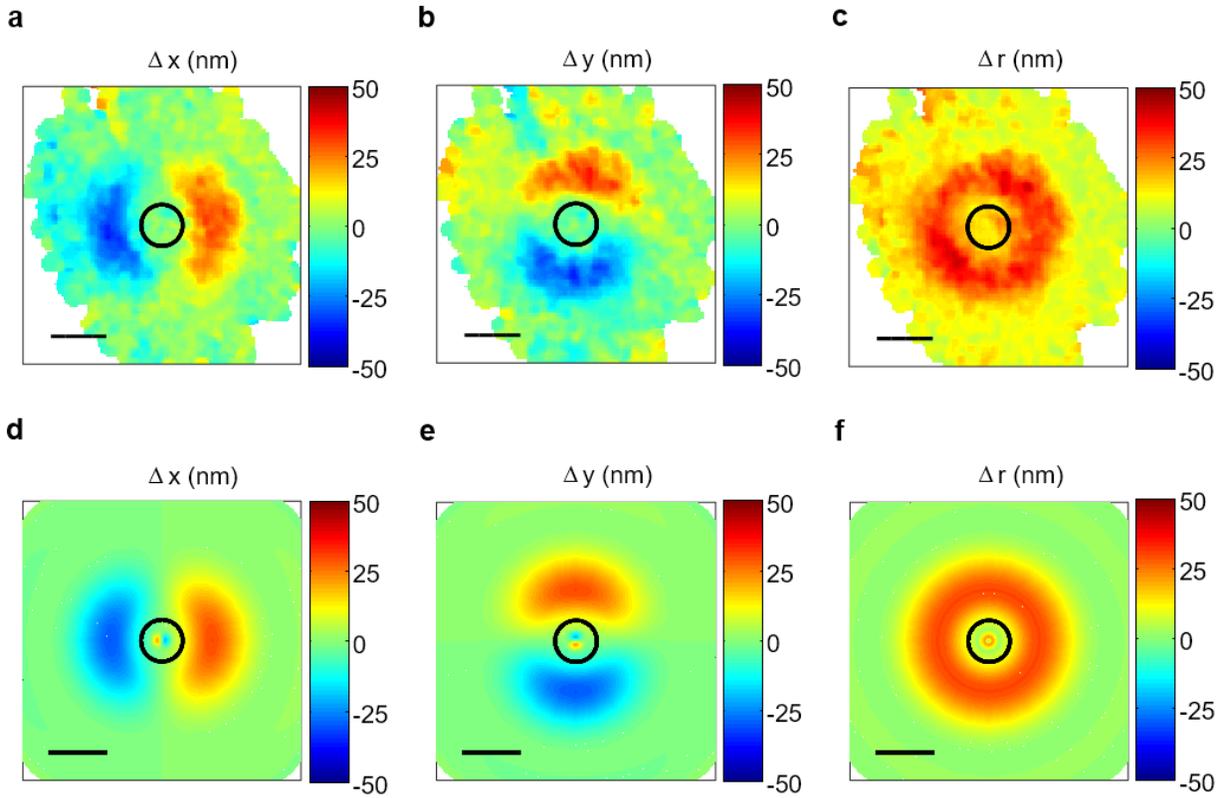

Figure 3. Images of centroid displacement coordinates $\Delta x$ (a), $\Delta y$ (b) and $\Delta r$ (c) as a function of quantum dot position relative to gold nanoparticle position. (d-f) FDTD simulation results corresponding to the measurements in panels (a-c) respectively. The black circles indicate the surface of a gold nanoparticle with a diameter of 150 nm. The scale bar is 200 nm.

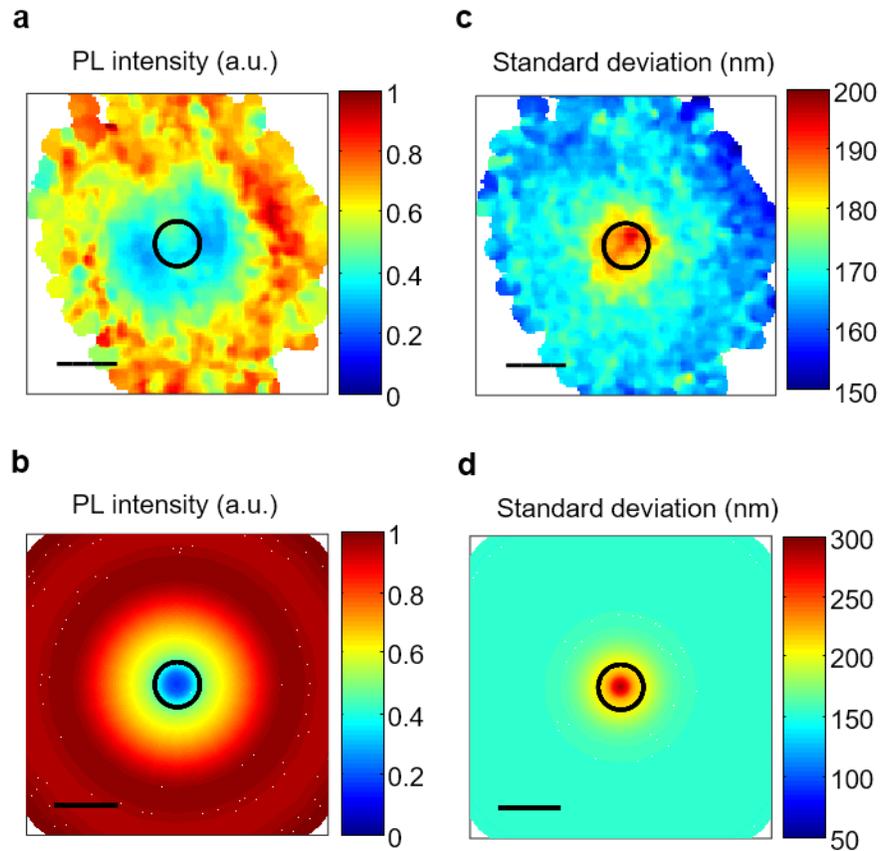

Figure 4. The measured (a) and calculated (b) emission intensity of the quantum dot as a function of its position relative to gold nanoparticle. Measured standard deviation of quantum dot diffraction spot (c) and calculated standard deviation (d) as a function of quantum dot position relative to gold nanoparticle position, respectively. The black lines indicate the surface of a gold nanoparticle with a diameter of 150 nm. The scale bar is 200 nm.